\documentclass[twocolumn,showpacs,preprintnumbers,amsmath,amssymb, prb]{revtex4}
\usepackage{graphicx}% Include figure files
\usepackage{dcolumn}% Align table columns on decimal point
\usepackage{bm}% bold math
\usepackage{wasysym}
\usepackage{textcomp}
\usepackage{epstopdf}
\usepackage{pxfonts}
\usepackage{txfonts}
\usepackage{tipa}
\usepackage{fontenc}
\usepackage{color,soul}
\usepackage{marvosym}
\begin{document}
\title{Superconductivity in non-centrosymmetric BiPd system}
\author{Bhanu Joshi, A. Thamizhavel and S. Ramakrishnan}
\address{Tata Institute of Fundamental Research, Mumbai-400005, India}
\begin{abstract}
\noindent
In this work, we establish the bulk superconductivity of  a high quality sample of monoclinic BiPd ($\alpha$-BiPd, space group P2$_1$) below 3.87~K by studying its electrical resistivity, magnetic susceptibility and heat capacity. We show that it is clean type-II superconductor with moderate electron-phonon coupling and determine its superconducitng and normal state parameters. 
Although $\alpha$-BiPd is a noncentrosymmetric superconductor with large electronic heat capacity (therefore, large $\gamma$), the effect of spin-orbit splitting of the electronic bands at the Fermi level
is small. This makes little influence on the superconducting properties of $\alpha$-BiPd.
\vskip 1truecm 
\noindent 
Ms number ~~~~~~~~~~~~PACS number:~72.10.Fk, 72.15.Qm, 75.20.Hr, 75.30.Mb\\
\end{abstract}
\maketitle
%\newpage
\section{Introduction}
\label{sec:INTRO}
\noindent
Ever since the discovery of the non-centrosymmetric heavy fermion superconductor CePt$_3$Si \cite{Bauer1}, there is widespread research activity to understand the nature of superconductivity  in such unconventional superconductors. The term ''non-centrosymmetric'' characterizes the symmetry of a crystal lattice without inversion center. In such materials the standard classification in even-parity spin-singlet and odd-parity spin triplet superconducting phases is obsolete, because the electrons are exposed to antisymmetric spin-orbit coupling, e.g. Rashba-type of coupling \cite{Rasbha} which arises due to electric field gradient in the crystal which has no inversion symmetry. An inherent feature is then the mixing of spin-singlet and spin-triplet Cooper pairing channels which are otherwise distinguished by parity. This mixing of pairing states is expected to cause a two-component order parameter. New forms of pairing appear giving rise to unusual temperature/field dependence of the superconducting parameters. During recent years we have witnessed rapid developments on the side of experiments and synthesis of novel materials as well as in the theoretical understanding of this type of superconductors. Indeed many new materials have been found, in particular, among the heavy fermions for which unconventional Cooper pairing is expected. Exotic mixed-state phases (vortex matter) have been predicted in theory and are under experimental investigations. For example in the case of CePt$_3$Si where superconductivity occurs at ambient pressure, whereas, transition to such a state happens in UIr \cite{UIr}, CeRhSi$_3$ \cite{Kimura} and CeIrSi$_3$\cite{Sugitani} only under pressure. However, the study of superconductivity in non-centrosymmetric materials which do not exhibit heavy fermion features is also important since it avoids additional complication that arises due strong f-electron correlations. Discovery of such materials also continues to increase, starting from binary carbides (R$_2$C$_{3-\delta}$ with R=La or Y) \cite{LaY}, 
Cd$_2$Re$_2$O$_7$\cite{CdRe} , Li$_2$(Pd,Pt)$_3$B \cite{Badi,Huan}, Mg$_2$Al$_3$ \cite{MgAl} and the recently found BaPtSi$_3$ \cite{Bauer2}. However, many of them (except Li$_2$Pt$_3$B) exhibit conventional BCS-like superconductivity due to small spin-orbit scattering. One of the common features in these compounds which show conventional superconductivity with small spin-orbit scattering is the absence of the large density of states at the Fermi level. Hence, it will be of interest to study a superconducting material which have conduction electrons with  high density of states at the Fermi level (not from high f-electron correlations) but has  no inversion symmetry. In this work we report our investigations in one such material, namely $\alpha$-BiPd, that shows bulk superconductivity below 3.87 K via, resistivity, magnetization and heat capacity studies. $\alpha$-BiPd has no inversion symmetry in its monoclinic crystal structure (space group P2$_1$). One of the important features of the structure is that the presence of short Pd-Pd bonds (shorter than those present in pure Pd metal). This could results in conduction electrons with large density of states at the Fermi level which is suggested in an earlier study \cite{BiPd1}. 
\noindent
\section{EXPERIMENTAL DETAILS}
\label{sec:EXPT}
\noindent
The compound BiPd undergoes  polymorphic transformation from $\alpha-$BiPd (monoclinic, P2$_1$) \cite{BiPd}  to $\beta$-BiPd (orthorhombic,Cmc2$_1$) above 210$^{\circ}$C. We have synthesized phase pure $\alpha$-BiPd which has a monoclinic structure with the space group P2$_1$ with b as its unique axis. Due to its low melting point ($~$650$^{\circ}$C), we have chosen to make the sample using a modified Bridgeman technique.
The sample was made by control heating the individual components (Bi, 99.999\% pure and Pd, 99.99\% pure) in a  high purity Alumina crucible with a pointed bottom which is kept in a quartz 
tube that is sealed under a vacuum of 10$^{-6}$ mbar. Initially  the contents were heated upto 650$^{\circ}$C (melting point of BiPd)  in 12 hrs and then kept at 650$^{\circ}$C for 12 hrs. Thereafter it was slow cooled to 
590$^{\circ}$C with  a rate of 1$^{\circ}$C/hr and finally the furnace was switched off. We obtained high quality poly and single crystals of few mm size with mass ranging from 10 to 50 mgm.  A piece of the from the melt 
was crushed into a fine powder for powder x-ray diffraction measurement using Cu K$\alpha$ radiation in a commercial diffractometer. The unit cell of the monoclinic structure ($\alpha$-BiPd, space group P2$_1$) is shown in Fig. 1.
\begin{figure}[h]
\includegraphics[width=0.45\textwidth]{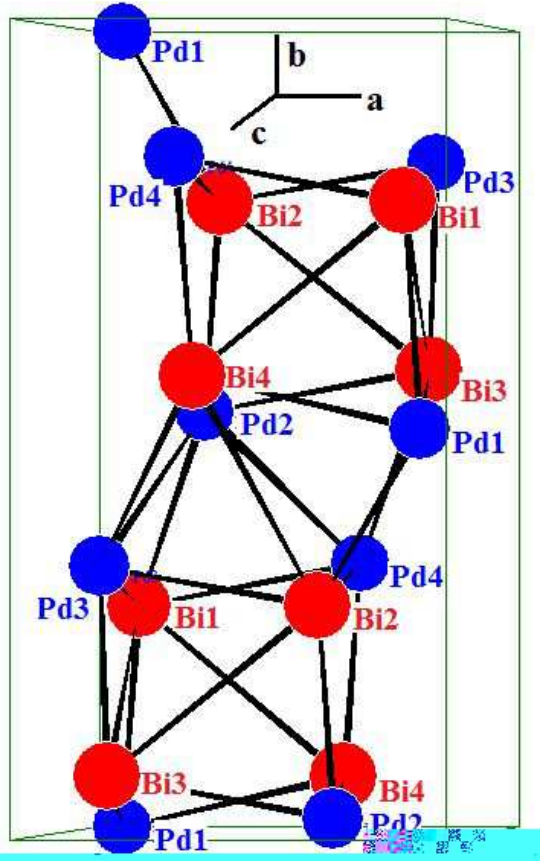}
\caption{\label{fig1}(Color online) $\alpha$-BiPd has Monoclinic(P2$_1$) structure with 16 atoms in a unit cell (8 formula unit). It contains four inequivalent Bi sites and  four inequivalent Pd sites.
It has no inversion symmetry in its monoclinic structure. This structure is stable below 200$^{\circ}$C.}
\end{figure}
 The structure consists four inequivalent sites for Bi and four inequivalent sites for Pd having 16 atoms in the unit cell. It also has alternate layer of Bi and Pd sheets with short Pd-Pd distances (shorter than those even in pure Pd metal) with no inversion symmetry in the structure. The Rietveld fit \cite{Riet} to the powder X-ray data is shown in Fig. 2. 
\begin{figure}[h]
\includegraphics[width=0.45\textwidth]{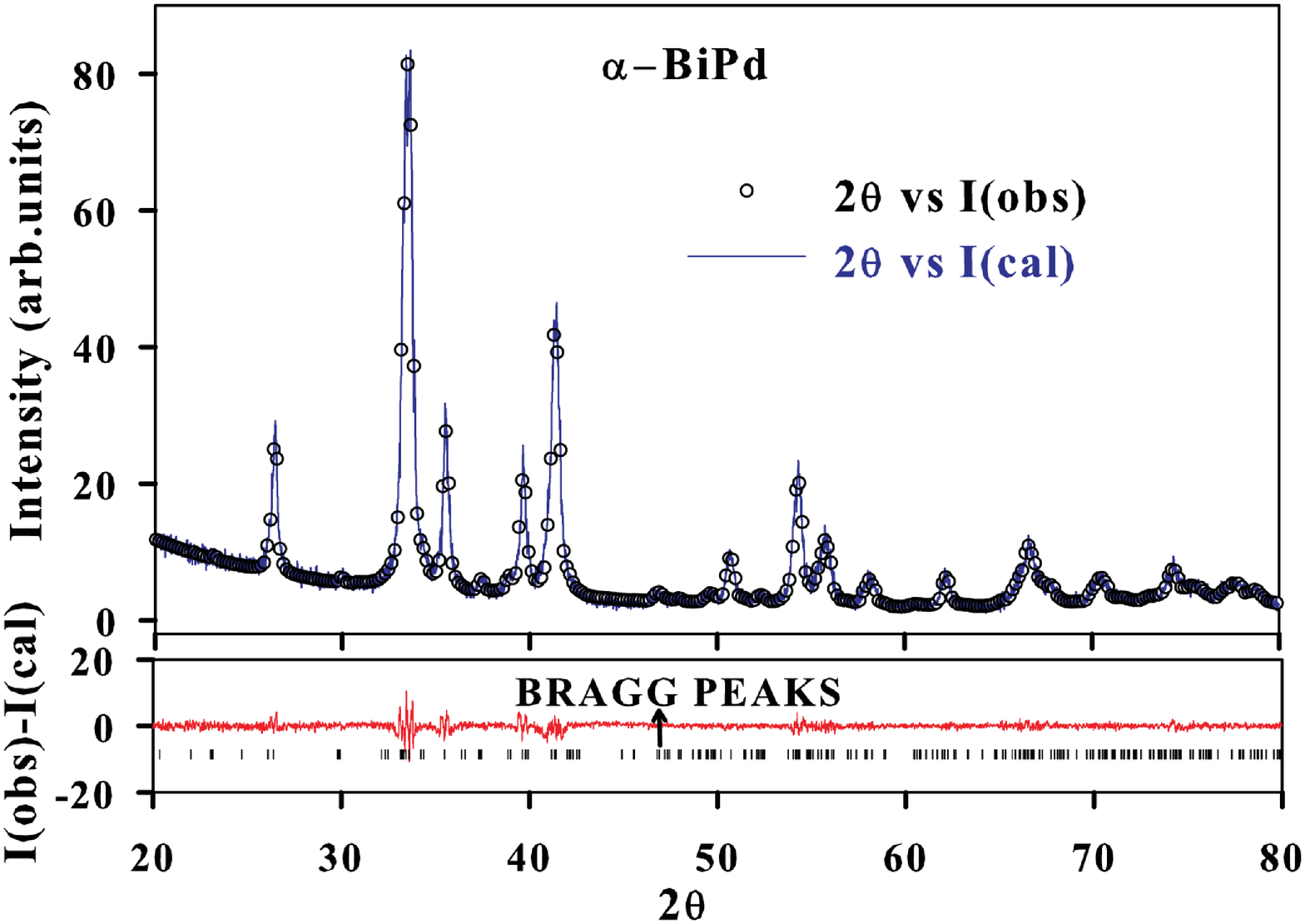}
\caption{\label{fig2}(Color online) Powder X-ray diffraction data of the monoclinic (P2$_1$) $\alpha$-BiPd.The solid line is the simulated data using FullProf(Rietveld program) with the unit cell
parameters obtained from Ref. 13.}
\end{figure}
The values for the lattice constants estimated from the fit are  a= 5.6284($\pm$0.0004)\AA, b= 10.6615($\pm$0.0004)\AA, 
c= 5.6752($\pm$0.0004)\AA, $\alpha$$=$$\gamma$$=$90 and $\beta$$=$101. These values are in agreement with an earlier report \cite{BiPd}  except for the value of $\beta$.\\
A commercial SQUID magnetometer (MPMS 5, Quantum Design, USA) was used to measure the temperature dependence of
the magnetic susceptibility $\chi$ in a field of 5 mT for temperatures between 1.8 to 20~K to detect the supercondcuting transition and in a field of 0.1~T in the temperature range from 10 to 300~K. The resistivity was measured using a four-probe ac technique on a home built setup with contacts made using silver paint on a bar shaped sample 1~mm thick, 10~mm long and 2 mm wide. The temperature was measured using a calibrated Si diode (Lake~Shore~Inc., USA) sensor. The sample resistance was measured with a  LR 700 AC Bridge (Linear Research, USA)  with  a current of 5~mA. The heat capacity was measured using a commercial setup (PPMS, Quantum Design, USA) in the temperature range from 2 to 200 K. 

\section{RESULTS AND DISCUSSION}
\label{sec:RD} 
\noindent
\subsection{Resistivity studies}
\label{sec:RES}
Fig.~3(c) shows the temperature dependence of the resistivity ($\rho$(T)) from 1.5 to 300 K. 
\begin{figure}[h]
\includegraphics[width=0.45\textwidth]{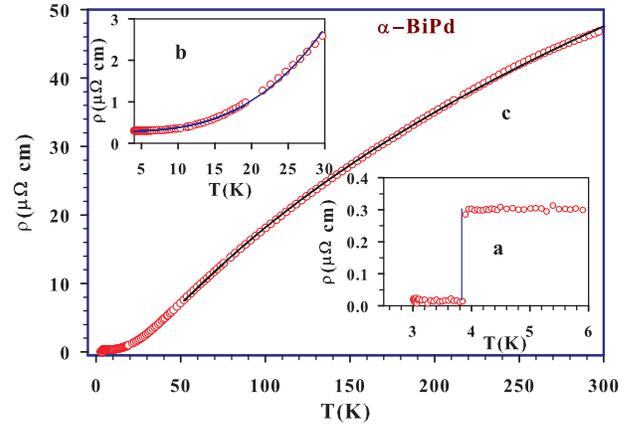}
\caption{\label{fig3}(Color online) Temperature dependence of the electrical resistivity of the monoclinic (P2$_1$) $\alpha$-BiPd. The insets a and b are described in the text.
The solid lines are fits to the equations described in the text.}
\end{figure}
The high quality of the sample is clearly evident from the large residual resistivity ratio 
($\rho_{300K}$/$\rho_{4K}$) of 160. The inset 3(a) shows the superconducting transition below 3.87 K  with a width of 10 mK. The inset 3(b) shows the low temperature resistivity from 4 to 30 K. The solid line is a fit to the equation,
$$ \rho (T)~=~\rho_0~+~A \times T^n \eqno(1) $$
The fit yields a value of 0.3 $\mu$$\Omega$~cm for $\rho_0$, 9.2$\times$ 10$^{-5}$ $\mu$$\Omega$~cm/K$^3$ for A and n=3. This T$^3$ dependence of the resistivity is interpreted by the Wilson's theory
\cite{Wilson} which takes into account the interband s-d phonon induced scattering within the Debye approximation. However, the same theory predicts a linear temperature dependence for the resistivity at higher temperatures (T$\geq$$\theta_D$, where $\theta_D$ is the Debye temperature) which is not observed in $\alpha$-BiPd (see Fig. 3(c)). The Wilson's theory neither considers the actual structure of the density of states of the electrons at the Fermi level nor the effect of unharmonicity of the phonon mode. But the contribution from these effects in the case of  $\alpha$-BiPd are small. There is yet another mechanism (parallel resistor model) suggested by Fisk and Webb \cite{Fisk} and later by Wisemann et al \cite{Wise} which accounts for the significant deviation of the resistivity from  the linear temperature dependence at high temperatures (100~K$<$T$<$300~K). This has been seen in  many other compounds where the $\rho$ value is rather high ($\approx$100 $\mu$$\Omega$ cm). The strong deviation from linearity and possible tendency towards saturation occur because the mean free path becomes short, of the order of few atomic spacings. When that happens, the scattering cross section will no longer be linear in the scattering perturbation. Since the dominant temperature-dependent scattering mechanism
is electron-phonon interaction here, the $\rho$ will no longer be proportional to the mean square atomic displacement, which is proportional to T for a
harmonic potential. Instead, the resistance will rise less rapidly than linearly in T and will show negative curvature (d$^2$$\rho$/dT$^2$~$<$0). This behavior is also seen in our previous  studies on silicides and germanides \cite{ramky1,ramky2,ramky3}. 

Wisemann {\it et al}  \cite{Wise}  describe the $\rho$(T) of these compounds (which is known as the parallel resistor model) where the expression of 
$\rho$(T) is given by, 
$$ {1\over{\rho(T)}} = {1\over{\rho_1(T)}} + {1\over{\rho_{max}}} ~, \eqno(2) $$ 
where $\rho_{max}$ is the saturation resistivity which is independent of temperature and $\rho_1$(T) is the ideal temperature-dependent resistivity. Further, the ideal resistivity is given by the following expression, 
 $$ \rho_1(T) = \rho_0 +C_1({T\over{\theta_D}})^3\int^{\theta_D/T}_0 {x^3dx\over{(1-exp(-x))(exp(x)-1)}} ~, \eqno(3) $$ where $\rho_0$ is the
residual resistivity and the second term is due to phonon-assisted electron scattering similar to the s-d scattering in transition metal compounds. $\theta_D$ is the Debye temperature and $C_1$ is a numerical constant. Eqn.(2) can be derived if we assume that the electron mean free path $l$ is replaced by $l+a$ ($a$ being an average interatomic spacing). Such an assumption is reasonable, since infinitely strong scattering can only reduce the electron mean free path to $a$. Chakraborty
and Allen \cite{Chakra} have made a detailed investigation of the effect of strong electron-phonon scattering within the framework of the Boltzmann
transport equation. They find that the interband scattering opens up new {\it nonclassical channels} which account for the parallel resistor model.
The high temperature resistivity fit (50~K$<$T$<$310~K) shown in the Fig. 3(c) yields a value of 164~$\mu$$\Omega$~cm for $\rho_{max}$, 0.33~$\mu$$\Omega$~cm for $\rho_0$, 76.3~$\mu$$\Omega$~cm/K$^3$
for C$_p$ and 168.6~K for $\theta_D$. This value to $\theta_D$ is close to the value obtained from the heat capacity data (described later) which suggests that the parallel resistor model can  successfully
explain the high temperature dependence of $\rho$(T) of $\alpha$-BiPd.
\subsection{Magnetic susceptibility studies}
\label{sec:SUS}
The temperature dependence of the magnetic susceptibility ($\chi$(T)) of $\alpha$-BiPd is shown in Fig. 4. 
\begin{figure}[h]
\includegraphics[width=0.45\textwidth]{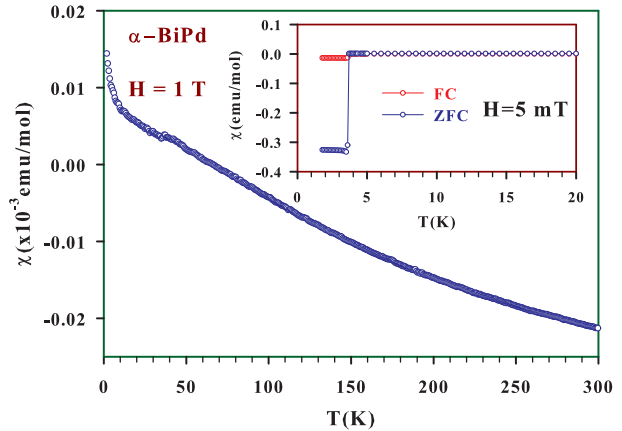}
\caption{\label{fig4} (Color online) Temperature dependence of the magnetic susceptibility of the monoclinic (P2$_1$) $\alpha$-BiPd in 1~T. The inset shows the temperature dependence of the  
susceptibility in a field of 5 mT in the zero field (ZFC) and field cooled (FC) states.}
\end{figure}
The inset shows the low temperature susceptibility data in 5 mT which reveals the diamagnetic transition  of $\alpha$-BiPd below 3.8~K that is in agreement with the resistivity data. The inset also shows the zerofield cooled and field cooled susceptibility data which elucidates low but significant pinning of the vortices in this compound. The high temperature $\chi$ data of $\alpha$-BiPd in its normal state is diamagnetic and its value ranges from -2.2$\times$10$^{-5}$ emu/mol at 300 K to 1.4$\times$10$^{-5}$ emu/mol at 1.8 K. The diamagnetic susceptibility at 300 K  is indeed surprising  given that the value of the
Sommerfeld coefficient ($\gamma$=41 mJ/mole K$^2$) is quite large. We believe that the observed diamagnetism in $\alpha$-BiPd is due to the large core diamagnetism of Bi.
In general, the observed value of the susceptibility can be written as,
$$ \chi_{obs}~=~\chi_{core}~+~\chi_{Landau}~+~\chi_{Pauli} \eqno(4) $$
$\chi_{core}$ is the core diamagnetism, $\chi_{Landau}$ is the Landau diamagnetism and  $\chi_{Pauli}$ is the Pauli paramagnetism. The equation (6) can be rewritten as,
$$ \chi_{obs}~-~\chi_{core}~=~\chi_{Pauli} [1-({1 \over 3}~({m \over m_{b}})] \eqno(5) $$
Here,  $\chi_{Pauli}$ = N$_A$~$\mu_B^2$~S~N(E$_F$) where N$_A$ is the Avogadro number, $\mu_B$ is the Bohr magneton, S is the Stoner factor, N(E$_F$) is the density of states at the
Fermi level, m is the free electron mass and m$_b$ is the band mass. The estimated value of $\chi_{Pauli}$ is 1.3$\times$10$^{-4}$ emu/mol. If one assumes the core diamagnetic susceptibility of Bi as -5$\times$10$^{-5}$ emu/mol and that of Pd as 0.2$\times$10$^{-6}$ emu/mol, we get a value -5.02 $\times$10$^{-5}$ emu/mol as the total contribution to the core diamagnetism in $\alpha$-BiPd.  
From this we calculate the Stoner enhancement factor S as 3. A  weak temperature dependence of $\chi$(T)  could arise due to two reasons. One of them could be the presence of magnetic  impurities at the ~100 ppm level in the compound. This could account for the observed small temperature dependence of 
the $\chi$(T). However, we believe that  impurity concentration of this amount either  in Bi or in Pd is unlikely (given that the limit of the impurities both in Bi and Pd elements imposed by the manufacturer is $\leq$10 ppm). The second reason could be the temperature variation of the density of states at
the Fermi level which results in a temperature-dependent Pauli spin susceptibility as seen in some of the A-15 compounds. Knight-shift measurements would be useful to resolve this issue.
\subsection{Heat capacity studies}
\label{sec:HC}
\begin{figure}[h]
\includegraphics[width=0.45\textwidth]{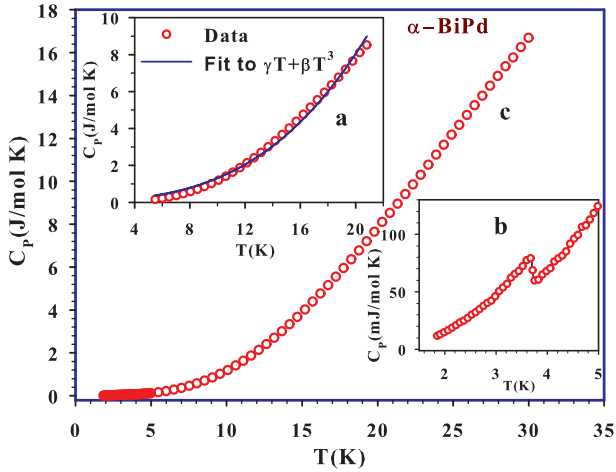}
\caption{\label{fig5}(Color online) Temperature dependence of the heat capacity of the monoclinic (P2$_1$) $\alpha$-BiPd from 2 to 35~K. The insets are explained in the text. Solid line is a fit to the equation(6)
described in the text.}
\end{figure}
The temperature dependence of the heat-capacity ($C_p$) from 2 to 30~K of $\alpha$-BiPd is shown in Fig. 5. The inset shows the low temperature 
C$_p$ vs T data. The jump in C$_p$ at 3.7~K ($\Delta$C= 30 mJ/mol K) clearly shows bulk superconducting ordering in this sample below this temperature. 
The temperature dependence of C$_p$ is fitted to the expression, 
$$ C_p~=~\gamma~T~+~\beta~T^3~, \eqno(6) $$ where $\gamma$ is due
to the electronic contribution and $\beta$ is due to the lattice contribution. The value of the ratio $\Delta$C$_p$/$\gamma$T$_c$ is 0.2 which is
significantly reduced from the BCS value of 1.43. Low values of $\Delta$C$_p$/$\gamma$T$_c$ have been observed before in the heat-capacity 
study of several superconducting compounds  \cite{ramky4,Vining}. According to these studies the reduced jump across T$_c$ could arise from extrinsic
effect (such as inhomogeneity in the sample or magnetic impurities) or from intrinsic effect (such as the existence of regions which do not
participate in superconductivity). In our sample of $\alpha$-BiPd, we estimate the impurity content to be less than 0.1 \% by volume and the
sharpness of the superconducting transition also suggests good homogeneity.It is possible that the reduced jump could arise from two-band
superconductivity where one band remains normal. However, detailed Fermi surface measurements are required before we can analyze the data in terms
of this model. The fit to the heat capacity data using the eqn.~6 in the temperature range from 5 to 20~K yielded 41~mJ/mol K$^2$ and 0.9 mJ/mol 
K$^4$ for $\gamma$ and $\beta$, respectively. The $\gamma$ value has been obtained by matching the entropy of normal and superconducting state at T$_c$ 
as suggested by Stewart {\it et al} \cite{Stewart}. From the $\beta$ value of 0.9 mJ/mol K$^4$, we estimate the $\theta_D$ to be 162~K using the relation,
$$ \theta_D~=~({12~\pi^4~N_A~n~k_B \over 5 \beta})^{1/3}~, \eqno(7) $$ where N$_A$ is the Avogadro's number, n is the number of atoms per formula
unit, and k$_B$ is the Boltzmann's constant.

\subsubsection{Upper critical field studies}
\noindent
The estimation of the upper critical field (H$_{c2}$) value at a given temperature has been made by measuring the resistance of the sample under a given magnetic field. The transition temperature in a given field is defined as the temperature which corresponds to the mid point of the resistance jump. The temperature dependence of 
H$_{c2}$ is shown in Fig. 6.
\begin{figure}[h]
\includegraphics[width=0.45\textwidth]{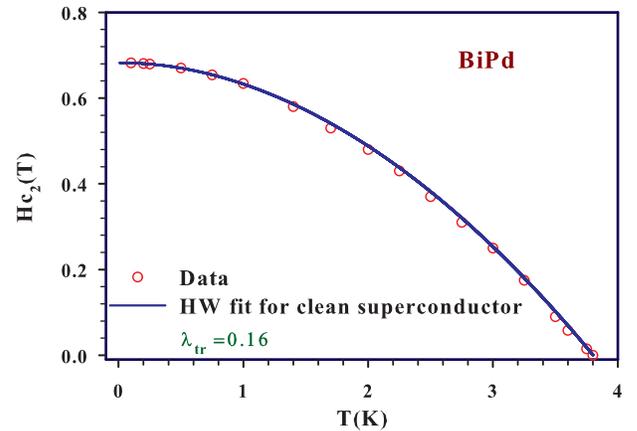}
\caption{\label{fig6} (Color online) Temperature dependence of the upper critical field (H$_{c_2}$(T)) of $\alpha$-BiPd. The estimation was made using both resistivity and magnetization measurements as
described in the text.}
\end{figure}
It is well known that in non-magnetic superconductors, the magnetic field interacts with the conduction electrons basically through two different mechanisms. Both lead to pair breaking and eventually destroys the superconductivity at a given field which is known as the critical field. One of this mechanisms arises due to the interaction of the field with the orbital motion of the electrons (orbital pair breaking) and the other is due to the interaction 
of the field with the electronic spin (Pauli paramagnetic limiting effects). Orbital pair breaking is the dominant mechanism at low fields and at very high fields, Pauli paramagnetic effect limits the upper critical field. We have fitted this temperature dependence of H$_{c2}$ to the theory of clean type-II superconductor  \cite{HW,Gor}. 
We obtain a value of  0.16 for $\lambda_{tr}$ , 0.69 T for H$_{c_2}$(0) and 3 kOe/K for dH$_{c2}$/dT near T$_c$. The value of the Pauli paramagnetic limiting field 
(H$_{Pauli}$=1.84$\times$T$_c$ in (Tesla)) for $\alpha$-BiPd  is very large (7.05 T) compared to the estimated value of the upper critical field at 0~K. 
From the GLAG \cite{GLAG1,GLAG2,GLAG3} theory, we get the value of the upper critical field at 0~K as,
$$ H_{c_2}(0)~=~0.72~(dH_{c_2}/dT)~T_c \eqno(10) $$
From this we get a value of 0.8~T for $ H_{c_2}(0)$ which is larger than the extrapolated experimental value of H$_{c_2}$. However, assuming the clean limit for the type-II superconductor, 
one can also estimate the dH$_{c2}$/dT using the relation,
$$ dH_{c_2}/dT~=~9.55~10^{24}~\gamma^2~T_c~[n^{2/3}~(S/S_F)]^{-2} (in~Oe/K) \eqno(11) $$
where $\gamma$ is the electronic heat capacity coefficient (ergs/cm$^{3}$K$^{2}$), n is the conduction electron density in units of cm$^{-3}$ and S/S$_F$
is the ratio of the area of the Fermi surface to that of a free electron gas of density n. Substituting the values of 
$\gamma$,n and assuming a simple model of spherical Fermi surface (S/S$_F$=1), we get a value 2.08 kOe/K which is  larger the value (3 kOe/K) obtained from the experiment. The reason
for this large discrepancy is not understood at this moment though similar anomalies in the value of dH$_{c2}$/dT have been reported in  earlier 
studies \cite{shelton1,shelton2}. In those earlier reports they have studied strong-coupled  superconductors which have complex phonon spectra. In that 
case, utilizing the theory to analyze is not strictly valid as the theory assumes electron interact via weak-coupling BCS type interaction 
potential and have spherical Fermi surface.
\subsection{Estimation of normal  and superconducting states parameters}
\label{sec:PARAM1}
\noindent
Using the value of $\theta_D$ and T$_c$, we can estimate the electron-phonon scattering parameter, $\lambda$, from the McMillan's theory \cite{McMillan} where $\lambda$ is given by,
$$ \lambda~=~{1.04~+~ \mu^*~ln(\theta_D/1.45~T_c) \over (1-0.62~\mu^*)~ln(\theta_D/1.45~T_c)-1.04} \eqno(12) $$
Assuming $\mu^*$=0.1, we find the value of $\lambda$ to be 0.72 which puts $\alpha$-BiPd as an intermediate coupling superconductor. 
On the basis of purely thermodynamical arguments, the thermodynamic critical field at T=0~K
(Hc(0)) can be determined by integrating the entropy difference between the superconducting and normal states. From our experimental heat capacity
data, we obtain a value of 2000~Oe for Hc(0). One can also calculate the thermodynamical critical field Hc(0) from the expression \cite{Orlando},
$$ H_c(0)~=~4.23~\gamma^{1/2}~T_c~, \eqno(13) $$ 
where $\gamma$ is the heat capacity coefficient(erg/cm$^3$ K$^2$).
This gives a value of Hc(0) as 2050~Oe. We can estimate the Ginzburg-Landau
coherence length $\xi_{GL}$ at T=0~K from the relation,
$$ \xi_{GL}(0)~=~ {5.87~10^{-17}~n^{2/3}~(S/S_F) \over \gamma~T_c}~, \eqno(14) $$
where $\gamma$ is electronic heat capacity coefficient (erg/ cm$^3$ K$^2$) and S/S$_F$ is the ratio of the area of the Fermi surface to that 
of a free electron gas of density n. This equation yields a value of 205~\AA~~for~$\xi_{GL}(0)$. If one uses the standard relation,
$ H_{c_2}(0)~=~~{\phi_0 \over 2~\pi~\xi_{GL}(0)^2} $ where $\phi_0$ is the flux quantum (2.07$\times$10$^{-7}$ oersted/cm$^2$,  we get a value 
of 218 \AA for $\xi_{GL}$(0) which is closer to  the estimated value from the equation (14). Using this value for  $\xi_{GL}$(0), one can estimate
$\kappa$(0) as 2.44 (since $ \kappa(0)~=~{H_{c_2}(0)\over \sqrt{2}~H_c(0)} $). However, from the GLAG theory for clean superconductors \cite{Orlando},
we know that $ \kappa(0)~=~{1.60~10^{24}~T_c~\gamma^{3/2}~\over [n^{2/3}~(S/S_F)]^2} $ (where $\kappa(0)$ = $\lambda_{GL}$(0)/$\xi_{GL}$(0). 
Using that, we get the $\kappa$(0) value as 2.7 which is close to the value estimated earlier. From the value of $\xi_{GL}$(0)~=~218~\AA, we get a value of 532~\AA~ for 
the Ginzburg-Landau penetration depth at 0 K ($\lambda_{GL}$(0)). The lower critical value can be determined by using the
relation, $$ Hc_1(0)~=~{Hc~(0)~ln[\kappa(0)]~\over 2^{1/2}~\kappa(0)}~, \eqno(15) $$
which yields a value of 516~Oe for the lower critical field at 0~K. This value of Hc$_1$(0)is in agreement with magnetization measurements on the
same sample.  The enhanced density of states can be calculated using the 
relation, $$ N^*(E_F)~=~0.2121~\gamma /N ~, \eqno(16) $$ where N is the 
number of atoms per formula unit and $\gamma$ is expressed in mJ/mol~K$^2$. 
The value of N$^*$(E$_F$) is 4.2~states/(eV~atom~spin-direction) and the value of the bare density of states N(E$_F$)~=~N$^*$(E$_F$)/(1+$\lambda$)
= 2.4 states/(eV~atom~spin-direction).  

The parameters are calculated using the Ginzburg-Landau theory for clean type-II superconductors. To verify the self consistency of our approach, 
we have estimated the mean free path ($l$) of our sample using the expression,
$$ l~=~ 1.27~10^4~[\rho~n^{2/3}~(S/S_F)]^{-1} \eqno(17)~,  $$
where n is the conduction electron density in units of cm$^{-3}$ and S/S$_F$
is the ratio of the area of the Fermi surface to that of a free electron gas of density n. If one assumes a simple model of spherical Fermi surface (S/S$_F$=1), the value of $l$ would be 1965 \AA~ and one can also calculate the value of the BCS coherence length (for S/S$_F$=1) from the expression,
$$ \xi_0~=~7.95~10^{-17}~[n^{2/3}]~(\gamma~T_c)^{-1}~, \eqno(18) $$ 
where $\gamma$ is expressed in ergs/cm$^3$ K$^2$. The value of $\xi_0$ was found to be 277 \AA~ which is much lower than $l$ which implies $\alpha$-BiPd is a clean type II superconductor. Moreover, we get the value
of 0.13 for $\lambda_{tr}$ using the equation $ \lambda_{tr}~=~{0.882~\xi_0 \over l} $, which is close to the value of 0.16 obtained from the HW fit earlier.
\subsection{Conclusion}
\label{sec:CON}
Noncentrosymmetry (NCS) introduces an electrical field gradient in the crystal which creates a Rashba-type antisymmetric spin-orbit coupling \cite{Rasbha}. An inherent feature is then the mixing of spin-singlet and spin-triplet Cooper pairing channels which are otherwise distinguished by parity. This mixing of pairing states is expected to cause a two-component order parameter. Such a behavior is observed in heavy fermion compounds like CePt$_3$Si \cite{Bauer1}, CeRhSi$_3$ \cite{Kimura}, CeIrSi$_3$ \cite{Sugitani} and UIr \cite{UIr}. Apart from heavy fermion properties, the occurrence of superconductivity in materials without inversion symmetry deserves its own merit. Therefore, we aim to find NCS superconducting materials without strong correlations among d or f electrons. This may set the stage for reliable electronic structure calculations, proving the splitting of bands due to missing inversion symmetry. Previous examples of superconductors without strong electron correlation, where, however, NCS was not a central issue of investigation, are binary  R$_2$C$_3$ with R=La or Y \cite{LaY} or Cd$_2$Re$_2$O$_7$ with Tc=1 K \cite{CdRe}. More recently the solid solution Li$_2$Pd,Pt$_3$B was demonstrated to show superconductivity \cite{Badi}. While Li$_2$Pd$_3$B is accounted for in terms of a conventional BCS-like superconductor, Li$_2$Pt$_3$B refers to unconventionality \cite{Huan} presumably due to a very large spin-orbit coupling. A recent study on Li$_2$Pd$_x$Pt$_{3-x}$B by muon-spin rotation and specific  heat, however, suggests that the whole series comprises single-gap s-wave superconductivity
\cite{Haef}. As far as $\alpha$-BiPd  is concerned, all experimental parameters and analyses point to a BCS-like superconductivity, which is based on spin-singlet pairing with a fully gapped DOS at the Fermi level. Therefore, $\alpha$-BiPd is classified as a type II superconductor in the clean limit. Here, the absence of inversion in the crystal structure does not give rise to anomalous features of the superconducting state such as a mixing of spin-singlet and spin-triplet pairs, as well as nodes in the superconducting gap at the Fermi surface. The overall effect of non-inversion symmetry seems to be of minor importance for the superconducting properties. In this respect, $\alpha$-BiPd seems to be in line with conclusions drawn for Li$_2$Pd$_3$B and BaPtSi$_3$ \cite{Bauer2}. Band structure calculations of $\alpha$-BiPd are in progress.

\end{document}